\documentclass[manuscript,screen]{acmart}

\usepackage{comment}
\usepackage{todonotes}

\AtBeginDocument{%
  \providecommand\BibTeX{{%
    \normalfont B\kern-0.5em{\scshape i\kern-0.25em b}\kern-0.8em\TeX}}}

   \setcopyright{iw3c2w3}
   \copyrightyear{2021}
   \acmYear{2021}
   \acmDOI{}

    \acmConference[IR4Children '21]{IR for Children 2000-2020: Where Are We Now? -- Workshop co-located with the 44$^{th}$ International ACM SIGIR Conference on Research and Development in Information Retrieval}{July 15, 2021}{Online Event}

\begin{document}

\title[Does your robot know?]{Does your robot know? Enhancing children's information retrieval through spoken conversation with responsible robots}

%%
%% The "author" command and its associated commands are used to define
%% the authors and their affiliations.
%% Of note is the shared affiliation of the first two authors, and the
%% "authornote" and "authornotemark" commands
%% used to denote shared contribution to the research.

%\authornote{Both authors contributed equally to this research.}

\author{Thomas Beelen}
\email{t.h.j.beelen@utwente.nl}
%\orcid{}
\affiliation{%
  \institution{University of Twente}
  \city{Enschede}
  \country{The Netherlands}
}

\author{Ella Velner}
\email{p.c.velner@utwente.nl}
\orcid{0002-9044-577X}
\affiliation{%
  \institution{University of Twente}
  \city{Enschede}
  \country{The Netherlands}
}

\author{Roeland Ordelman}
\email{r.j.f.ordelman@utwente.nl}
\orcid{}
\affiliation{%
  \institution{University of Twente}
  \city{Enschede}
  \country{The Netherlands}
}

\author{Khiet P. Truong}
\email{k.p.truong@utwente.nl}
\affiliation{%
  \institution{University of Twente}
  \city{Enschede}
  \country{The Netherlands}
}

\author{Vanessa Evers}
\email{vanessa.evers@ntu.edu.sg}
\affiliation{
 \institution{NTU Institute of Science and Technology for Humanity}
 \city{Singapore}
 \country{Singapore}
}

\author{Theo Huibers}
\email{t.w.c.huibers@utwente.nl}
\orcid{0002-9837-8639}
\affiliation{%
  \institution{University of Twente}
  \city{Enschede}
  \country{The Netherlands}
}

%%
%% By default, the full list of authors will be used in the page
%% headers. Often, this list is too long, and will overlap
%% other information printed in the page headers. This command allows
%% the author to define a more concise list
%% of authors' names for this purpose.
%\renewcommand{\shortauthors}{ et al.}

%%
%% The abstract is a short summary of the work to be presented in the
%% article.

% \url{https://www.fab4.science/IR4c/}

\begin{abstract}

In this paper, we identify challenges in children's current information retrieval process, and propose conversational robots as an opportunity to ease this process in a responsible way. Tools children currently use in this process, such as search engines on a computer or voice agents, do not always meet their specific needs. The conversational robot we propose maintains context, asks clarifying questions, and gives suggestions in order to better meet children's needs. 
Since children are often too trusting of robots, we propose to have the robot measure, monitor and adapt to the trust the child has in the robot. This way, we hope to induce a critical attitude with the children during their information retrieval process. %We work towards conversational robots that ease children's information retrieval process, whilst maintaining a responsible trust relationship. 

\end{abstract}
%%
%% The code below is generated by the tool at http://dl.acm.org/ccs.cfm.
%% Please copy and paste the code instead of the example below.
%%


% https://dl.acm.org/ccs 
%\ccsdesc[500]{Computer systems organization~Embedded systems}

%%
%% Keywords. The author(s) should pick words that accurately describe
%% the work being presented. Separate the keywords with commas.

% \keywords{datasets, neural networks, gaze detection, text tagging}

%% A "teaser" image appears between the author and affiliation
%% information and the body of the document, and typically spans the
%% page.
\begin{comment}
\begin{teaserfigure}
  \includegraphics[width=\textwidth]{sampleteaser}
  \caption{Seattle Mariners at Spring Training, 2010.}
  \Description{Enjoying the baseball game from the third-base
  seats. Ichiro Suzuki preparing to bat.}
  \label{fig:teaser}
\end{teaserfigure}
\end{comment}

%%
%% This command processes the author and affiliation and title
%% information and builds the first part of the formatted document.
\maketitle

\section{Introduction}

The UN convention of children's rights article 17 states that children have the right to access to information in a way that they can understand \cite{unicef}. However, children still struggle to access information with the tools they currently use, such as search engines and voice agents (e.g. \cite{druin_how_2009, lovato_hey_2019, jochmann-mannak_children_2010}). Current search engines do not always support the way children formulate search queries \cite{druin_how_2009}. Studies suggest that children have a more limited knowledge base and vocabulary to recall search terms from \cite{hutchinson_how_2005}. Additionally, spelling and typing issues hamper children's searches \cite{druin_how_2009, jochmann-mannak_children_2010}. It is observed that children can benefit from query suggestions in search engines they typically use \cite{jochmann-mannak_children_2010}. However, these are not tailored to children \cite{fails_query_2019}. While voice agents have become available, which children also use to search for information \cite{lovato_hey_2019, lovato_siri_2015, lopatovska_talk_2019, garg_he_2020}, the interaction that these voice agents provide is of a simple question-answer style, and not conversational. This limits children in that they cannot ask follow up questions like they often assume \cite{lovato_hey_2019, landoni_sonny_2019}, have to include context into complex statements \cite{lovato_hey_2019, yarosh_children_2018}, and do not get assisted with suggestions or clarifying questions \cite{druga_hey_2017, yarosh_children_2018}.

Inspired by \citet{landoni_sonny_2019}, we propose to aid children in searching and exploring (multimedia) information through \emph{spoken conversational agents}. Our target group is children of 10-12 years old. Children in these ages are starting to develop logical reasoning, but still have limited skills in reading and writing \cite{gossen_specifics_2013}. Through spoken conversation, the agent can support the child in formulating their information needs in a collaborative way by asking questions and proposing suggestions. To facilitate an engaging interaction \cite{li_benefit_2015}, a \emph{physically embodied} agent\footnote{from now on simply referred to as \emph{robot}} will be used. While embodied agents afford rich ways (e.g., visual, auditory) to design and tailor the interaction to children, they also give rise to challenges related to trust and ethical aspects that ask for responsible design. Children are prone to trust robots (too much) \cite{di_dio_shall_2020}, not only technologically, as they do with computers, but also interpersonally due to the social bond they create \cite{VanStraten2018TechnologicalStudy}. This might lead to children relying on false information. It would be beneficial to facilitate a critical attitude with the children during the interaction.

Our proposed research for the coming years has two objectives. We will investigate how to improve the information retrieval (IR) process for children with embodied conversational robots, while taking into account the child's trust in the robot and the information it provides (elaborated on in section \ref{sec:trust}). This brings us to the following research questions: 

\begin{enumerate}
\item How can conversational robots help children explore and search multimedia information through a child-tailored interaction?
\item How can conversational robots induce a critical attitude with children towards the robot and the information it provides during the search interaction?

\end{enumerate}

\section{Robots can help children find information}

In the introduction we described the issues children face when searching with search engines via a computer or voice agent. We envision a robot solution that addresses these issues by supporting a mixed-initiative, spoken, conversational interaction. This type of search conversation is described in the Spoken Conversational Search (SCS) paradigm \cite{trippas_towards_2020}. Further exploration of SCS for children is needed, since, to our knowledge, it has not been studied how SCS can be tailored to children. For instance, (when) do children get dissatisfied with clarifying questions (see \cite{wang_controlling_2021})?

In order to develop an agent that helps children communicate their information need, several challenges need to be addressed. Firstly, the envisioned conversational robot can maintain context, meaning it uses memory to store previous statements (see \cite{radlinski_theoretical_2017}). This makes it easier for children to communicate complex information needs. In a non-conversational interaction children need to integrate context into a single query which is difficult \cite{yarosh_children_2018}. As \citet{yarosh_children_2018} suggest, (embodied) agents could become more usable to children if they support context statements to be specified upfront, or as separate statements during the conversation. Maintaining context additionally enables children to ask follow-up questions, which children expect of voice agents during the search process \cite{landoni_sonny_2019, lovato_hey_2019, vtyurina_exploring_2017}. Secondly, during the conversational interaction, the robot can take initiative to improve the query by asking clarifying questions \cite{aliannejadi_asking_2019, druga_hey_2017}. These questions can address specific missing pieces of information \cite{yarosh_children_2018}, or simply clarify that the child was heard correctly. Thirdly, we will study how and when the robot can provide query suggestions in this interaction. These suggestions should be adapted to children's vocabulary and interests \cite{fails_query_2019}. By addressing these challenges we develop the envisioned mixed-initiative, spoken, conversational, robot. As a final challenge, we want to find out how this search process interaction works for children, and whether it indeed improves their IR process (for example using the framework by \citet{landoni_sonny_2019}).

\section{Children cannot always trust a robot} \label{sec:trust}
In addition to developing the child-tailored search interaction, it is important that this interaction is responsible. Children tend to trust robots too much, which is also known as overtrust \cite{di_dio_shall_2020}. Since children build social bonds with robots, more so than with computers, this trust is interpersonal as well as technological \cite{belpaeme_multimodal_2013, VanStraten2018TechnologicalStudy}. This trust relationship might lead to dangerous situations, since there is so much false information on the internet. Undertrust may lead to children discarding useful information. We mainly focus on overtrust, since undertrust leads to looking for the information elsewhere, while overtrust might lead children to rely on false information. It is therefore crucial that children remain critical to the information that the robot provides. This calls for the ability to measure, monitor, and lower the trust that the child has in the robot when appropriate. 

Currently, trust in child-robot interaction is mainly measured with questionnaires or trust games that do not fit the IR interaction, hence a new kind of measure needs to be developed \cite{Velner2021Traits}. Questionnaires need to be conducted after the interaction, which does not fit a real-world setting. Trust games need to be implemented within the interaction, which constricts the type of interaction. We envision the robot to be able to measure trust real-time during the interaction and to alter its behavior appropriately when trust is too high. Since the interaction will mainly be done via speech, and a lot of information on emotions and attitudes can be extracted from speech \cite{Fernandez2005ClassicalSpeech, Kim2017TowardsLearning, Scissors2008LinguisticCMC, TerMaat2010HowAgent}, we explore speech cues (e.g., acoustic features such as pitch and loudness, nonverbal vocalisations such as filled pauses and laughter, dialogue acts) as a measure of trust that the robot could use during the interaction. 

Having measured the current trust level, the robot should act on it. In case of overtrust the robot should display behaviors associated with a lower trust level. For example, when the robot gives an answer to the child about earth and what it looks like, and it notices overtrust, the robot could express uncertainty, to spark a critical attitude with the children. Such cues can be modeled after human interaction when a speaker presents information they are uncertain about \cite{Swerts2005AudiovisualKnowing}. 
Previous studies on children and trust have shown that children take into account past performance, and how the robot is introduced \cite{lane_informants_2013, geiskkovitch_what_2019}. Children can also base their trust on superficial cues, e.g., pretty or nice robots are perceived as more trustworthy \cite{bascandziev_beauty_2014,landrum_when_2013}. However, not all of these cues are suitable for real-time trust adaptation, e.g., a change in appearance during the interaction would not make sense. Furthermore, having robots make intentional mistakes could be considered unethical \cite{Katsanis2020CriteriaAutism}. Hence, it is still an open question as to what strategies the robot could use to lower a child's trust, and consequently induce a critical attitude towards the robot and the information it provides. 

Finally, it is not a given that children who trust a robot also trust the information that the robot provides, and vice versa. It is unclear how the trust in the robot affects the information acceptance. In one of our pilot studies, we introduced a child to a robot with untrustworthy characteristics, e.g., a neutral face, monotonous voice and wrong answers (please note that this is not suited for real-world applications, but still implemented here, because we need to study the workings of low and high trust in robots with children in order to lower the trust when appropriate), and asked them afterwards whether they would comply with the robot's answers in a quiz. This pilot was part of a larger study currently being conducted, which investigates the trust relationship between children and robots, and whether this relationship influences what they do with the information that the robot provides. One child indicated that they did not trust the robot, but still said \emph{``if the robot says so, it must be true''} when asked whether they would comply with the robot. While this could be illustrative of the people-pleasing nature of children during questionnaires and interviews, it calls for a more in depth investigation on the relation between the trust in a robot and the information it provides. For instance, children's behavior might reflect their trust better than their answers on a trust questionnaire. 

\section{Conclusion}
In this paper, we propose spoken conversational robots to improve children's information retrieval process. We identified challenges and opportunities in developing the conversational interaction in a way that children are supported in expressing their information needs. Some guidelines and solutions for developing spoken conversational search for adults already exist, however, for children, this is a rather new and underresearched territory. Moreover, since we are developing for children, extra attention should be paid to making the interaction not only child-tailored, fun and engaging, but also responsible, in terms of monitoring for overtrust. Introducing spoken conversational robots creates a tension between on the one hand the envisioned benefits of the conversational interaction, and on the other hand, the call for responsible design due to the emerging trust relationship between robot and child, and the effects this relationship can have on the information perception. We are excited to start addressing this multifaceted challenge in the coming years by a multidisciplinary team with experts in design, ethics, psychology, child-robot interaction, search technology, and computer science. 

\section{Acknowledgements}
This research is supported by the Dutch SIDN fund
(https://www.sidn.nl/) and TKI CLICKNL funding of the Dutch Ministry of
Economic Affairs (https://www.clicknl.nl/).

\bibliographystyle{ACM-Reference-Format}
\bibliography{references_combined, ella, references}

\end{document}